\documentclass[
aip,
pop,
amsmath,amssymb,
 reprint,%
]{revtex4-1}

\usepackage{graphicx}
\usepackage{dcolumn}
\usepackage{bm}

\usepackage[utf8]{inputenc}
\usepackage[T1]{fontenc}
\usepackage{mathptmx}

\usepackage{color}
\usepackage{amssymb}
\usepackage{natbib}
\usepackage{hyperref}
\usepackage{xcolor}

\begin{document}


\title[]{Effects of wave damping and finite perpendicular scale on three-dimensional Alfv\'en wave parametric decay in low-beta plasmas}

\author{Feiyu Li}
    \email{fyli.acad@gmail.com}
	\affiliation{New Mexico Consortium, Los Alamos, NM 87544, USA}
\author{Xiangrong Fu}
    \affiliation{Los Alamos National Laboratory, Los Alamos, NM 87545, USA}
    \affiliation{New Mexico Consortium, Los Alamos, NM 87544, USA}
\author{Seth Dorfman}
	\affiliation{Space Science Institute, Boulder, CO 80301, USA}
	\affiliation{University of California Los Angeles, Los Angeles, CA 90095, USA}

\date{\today}

\begin{abstract}
Shear Alfv\'en wave parametric decay instability (PDI) provides a potential path toward significant wave dissipation and plasma heating. However, fundamental questions regarding how PDI is excited in a realistic three-dimensional (3D) open system and how critically the finite perpendicular wave scale---as found in both laboratory and space plasmas---affects the excitation remain poorly understood.  
Here, we present the first 3D, open-boundary, hybrid kinetic-fluid simulations of kinetic Alfv\'en wave PDI in low-beta plasmas. 
Key findings are that the PDI excitation is strongly limited by the wave damping present, including electron-ion collisional damping (represented by a constant resistivity) and geometrical attenuation associated with the finite-scale Alfv\'en wave, and ion Landau damping of the child acoustic wave. The perpendicular wave scale alone, however, plays no discernible role: waves of different perpendicular scales exhibit similar instability growth as long as the magnitude of the parallel ponderomotive force remains unchanged. These findings are corroborated by theoretical analysis and estimates. 
The new understanding of 3D kinetic Alfv\'en wave PDI physics is essential for laboratory study of the basic plasma process and may also help evaluate the relevance/role of PDI in low-beta space plasmas. 
\end{abstract}

\maketitle

\section{Introduction} \label{introduction}

Alfv\'en waves represent a fundamental magnetohydrodynamic (MHD) mode with far-reaching implications for laboratory, space, and astrophysical plasmas. The interaction of Alfv\'en waves with energetic particles is crucial to the performance of burning fusion plasmas~\cite{chen2016physics}. Shear Alfv\'en waves are also an excellent carrier of significant magnetic and kinetic energy over large distances in space plasmas. Nonlinear processes associated with large-amplitude Alfv\'en waves are key to understanding several major problems such as turbulent cascades and plasma energization. As a prominent example, parametric instabilities are thought to potentially contribute to solar coronal heating~\cite{del2002coronal}, the observed spectrum and cross-helicity of solar wind turbulence~\cite{inhester1990drift,del2001parametric,yoon2008parallel}, and damping of fast magnetosonic waves in fusion plasmas~\cite{lee1998internal,oosako2009parametric}. In particular, the parametric decay instability (PDI)~\cite{sagdeev1969nonlinear,derby1978modulational,goldstein1978instability}, well established in theory for over half a century~\cite{sagdeev1969nonlinear,hasegawa1976parametric,derby1978modulational,goldstein1978instability,wong1986parametric,longtin1986modulation,hollweg1993modulational,hollweg1994beat}, produces a forward propagating ion acoustic wave (or sound wave which we use interchangeably hereafter) and a backward propagating Alfv\'en wave; this process may directly cause plasma heating and cascades of wave decays~\cite{chandran2018parametric,kiyani2015dissipation}. Theory also suggests a modulational instability, which results in forward propagating upper and lower Alfv\'enic sidebands as well as a non-resonant acoustic mode at the sideband separation frequency~\cite{hollweg1994beat}. 

Observational evidence of Alfv\'en wave PDI in space plasma has been reported. A satellite measurement in the ion foreshock region found a number of possible PDI events, yet the results were inconclusive as the “decay line” signatures were missing in many intervals~\cite{spangler1997observations}. An analysis of WIND spacecraft data suggested that the fluctuations of magnetic field and plasma density in the solar wind at 1 AU may be limited by the PDI~\cite{bowen2018density}. Hahn et al. recently reported an observational evidence of PDI in the lower solar atmosphere using remote measurements of spectral lines~\cite{hahn2022evidence}. However, space observations can be limited (e.g. by a turbulent environment and the lack of control over the process) and especially challenging in the near-Sun low-beta region, where the PDI is predicted to have largest growth rates.

To fully elucidate the relevance and role of PDI in space plasma dynamics, more controlled studies using either laboratory experiments or numerical modeling are needed. 
Such investigations under dimensionless and scaled parameters similar to that of some space plasma regions~\cite{howes2018laboratory,lichko2023enabling,dorfman2023nextgen} will help validate PDI theories and gain new insights into the spatiotemporal behavior and consequences of this basic plasma wave phenomenon. 
Experimental progress has been made over the past decade in studying PDI-related physics with the Large Plasma Device (LAPD), a flagship device hosted at UCLA uniquely suited for studying space-relevant Alfv\'en waves in low-beta plasmas~\cite{gekelman1997laboratory,gekelman1999review,maggs2003laboratory,vincena2006drift,auerbach2010control,auerbach2011resonant,howes2012toward,dorfman2013nonlinear,dorfman2016observation}. Using two counter-propagating Alfv\'en waves of comparable amplitudes, the three-wave coupling at the heart of PDI was verified by measuring a clear resonant peak in the acoustic beat wave response.~\cite{dorfman2013nonlinear}; more recently, PDI growth rates have been inferred from a reduction in the damping of a small-amplitude, counter-propagating seed wave when a large-amplitude pump wave is turned on~\cite{dorfman2023measuring}. 
Nevertheless, these experiments have thus far not been able to produce PDI in its standard form driven by a single Alfv\'en wave. When using a single wave driver, Alfvén wave sidebands and a low frequency nonresonant mode were produced~\cite{dorfman2016observation}. However, the spatial pattern of the child modes does not match predictions for the standard perpendicular wavenumber $k_\perp=0$ modulational instability, suggesting that perpendicular nonlinear forces play a key role in the observations. The standard $k_\perp=0$ modulational instability is also predicted to have a significantly smaller growth rate than PDI under the chosen set of experimental parameters~\cite{hollweg1994beat}.

On the other hand, extensive numerical modeling of PDI has been conducted, using either MHD simulations~\cite{del2001parametric,shi2017parametric}, hybrid simulations~\cite{terasawa1986decay,vasquez1995simulation,araneda2007collisionless,matteini2010parametric,fu2018parametric,gonzalez2020role,li2022parametric,li2022hybrid}, or even full particle simulations~\cite{nariyuki2008parametric,gonzalez2023particle}.
However, a periodic infinite system has been routinely adopted by many of these studies, lacking direct relevance to both the laboratory and space plasma settings which feature an open system with wave injection. Preliminary open-boundary simulations showed distinct energy transfer and partition from usual periodic boundary interactions~\cite{li2022parametric}.  
More critically, most of these simulations have focused on investigating the consequences of PDI, without addressing what conditions are needed to excite PDI in the first place. This problem is nontrivial as partly illustrated by the difficulty in demonstrating PDI in the laboratory, and directly determines the relevance of PDI in space plasmas.  

Toward addressing the excitation problem, we have recently developed  quasi-1D open-boundary hybrid simulations focusing on LAPD-relevant conditions~\cite{li2022hybrid}, and found the threshold amplitudes and frequencies of a planar zero-$k_\perp$ Alfv\'en wave required for exciting PDI under given plasma parameters. 
Physically, these thresholds were obtained by requiring PDI to grow faster than Landau damping of the acoustic mode, as well as the convective motion of both child modes in a bounded plasma. 
While the result is of interest to both the laboratory and space study at large perpendicular wave scales (i.e. the spatial extent across the background magnetic field), Alfv{\'e}n waves in both contexts can also develop significant wave $k_\perp$. In the low-beta solar coronal region, large $k_\perp$ may be induced by transverse plasma gradients, resonance absorption, and turbulent cascade~\cite{ofman1995alfven,hollweg1999kinetic,kiyani2015dissipation}. In the laboratory, $k_\perp d_i > 1$ ($d_i$ is the ion inertial length) due to the finite perpendicular antenna size necessary to fit the wave in the laboratory plasma column~\cite{morales1997structure}.   
These finite-$k_\perp$ kinetic Alfv\'en waves 
carry significant parallel electric current channels~\cite{hollweg1999kinetic,gigliotti2009generation}, fundamentally different from the plane-wave scenario~\cite{li2022hybrid}.
How the PDI excitation may be modified by the new 3D features remains poorly understood. 
Furthermore, previous PDI theories were mostly derived for $k_\perp=0$ plane waves~\cite{derby1978modulational,goldstein1978instability,hollweg1994beat}, although some theoretical/numerical studies allowed for child/parent waves with finite $k_\perp$~\cite{vinas1991parametric,matteini2010parametric,shi2019three}. It is not clear, and no present theory explores, how $k_\perp$ associated with a finite-perpendicular-scale pump wave may influence PDI development. 

In the present work, we present for the first time 3D open-boundary hybrid simulations of PDI driven by a single Alfv\'en wave of finite perpendicular scale.   
Our central new results are that the PDI excitation is found to be strongly limited by 3D wave damping of the child modes, including both the Alfv\'en wave damping and acoustic wave damping. Currently considered by the simulations are i) electron-ion collisional damping (represented by a constant resistivity $\eta$~\cite{winske1991hybrid}) and geometrical attenuation associated with the finite-frequency, finite-source-size Alfv\'en wave~\cite{morales1997structure}, and ii) ion Landau damping of the child acoustic wave. 
On the other hand, for a given magnitude of wave damping, the PDI excitation in a low-beta plasma is found to have no discernible dependence on $k_\perp$ alone, as long as the parallel ponderomotive force remains constant. This lack of a dependence on $k_\perp$ means that existing laboratory experiments, which can only produce highly oblique Alfv{\'e}n waves, may still be capable of demonstrating PDI excitation. 
In space plasmas, the effectiveness of PDI under large $k_\perp$ is important to establishing the relevance of PDI as wave energy cascades in the perpendicular direction towards a small dissipation scale.


\section{3D open-boundary hybrid simulation of PDI with a single finite-scale Alfv\'en wave} \label{setup}

\begin{figure*}[htp]
	\centering
	\includegraphics[width=0.9\textwidth]{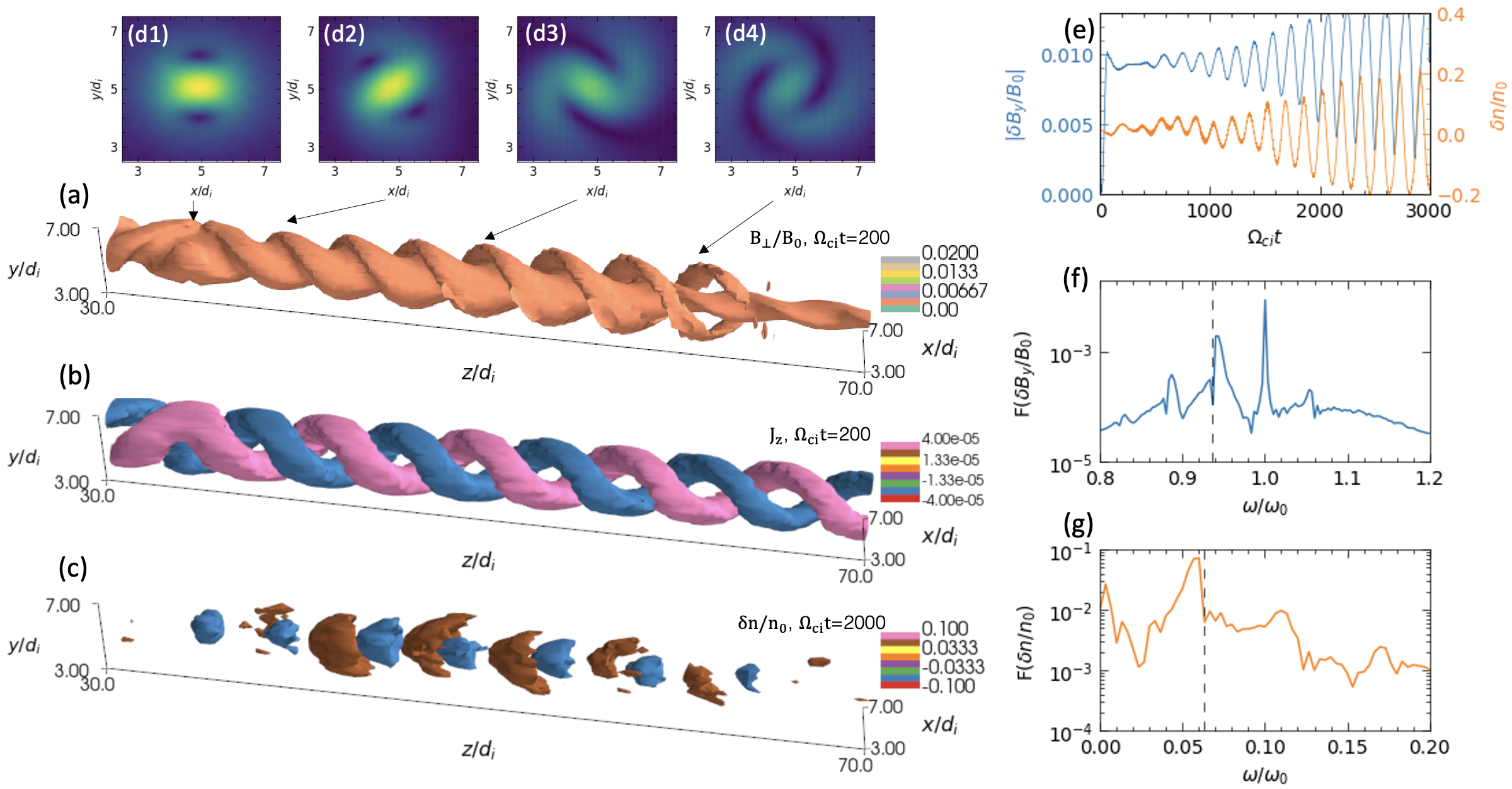}
	\caption{3D open-boundary hybrid simulation of PDI driven by a finite-scale, circularly polarized Alfv\'en wave. (a-c) 3D isosurfaces of the perpendicular wave magnetic field at $t=200\Omega_{ci}^{-1}$, parallel electric current density at $t=200\Omega_{ci}^{-1}$, and ion density fluctuations at $t=2000\Omega_{ci}^{-1}$, respectively. (d1-d4) The $xy$-cut of perpendicular magnetic field at $z=35, 40, 50, 60 d_i$, respectively, taken at $t=200\Omega_{ci}^{-1}$. (d1) corresponds to the plane at which the wave is injected. (e-g) The magnetic field envelope $|\delta B_y/B_0|$ and the density fluctuations probed at $(x,y,z)=(5,5,41) d_i$ and their corresponding Fourier spectra (obtained over the the full time window shown). In (e) the envelope of the field oscillation is shown, and its spectrum in (f) is performed on the fast field oscillations not shown in (e). The vertical dashed line in (f,g) refers, respectively, to the frequency of the child Alfv\'en wave and ion acoustic wave, as predicted from PDI theories~\cite{sagdeev1969nonlinear,derby1978modulational,goldstein1978instability}.}
	\label{fig1}
\end{figure*}

We start by introducing the 3D simulation setup (Fig.~\ref{fig1}) based on the H3D code~\cite{karimabadi2006global}, which models kinetic ions plus a massless electron fluid. The box/plasma occupies $z$ = [0,100]$d_i$ along the background magnetic field ($B_0$) direction, and two field masks (used for absorbing Alfv\'en waves~\cite{li2022parametric}) occupy $z$ = [0,30]$d_i$ and $z$ = [70,100]$d_i$. Only the central region  $z$ = [30, 70]$d_i$, containing actual Alfv\'en wave-plasma interactions, is displayed. The cell size along $z$ is $\Delta z=0.5d_i$. In this example, the perpendicular dimensions are of size $L_x=L_y=10 d_i$ and sampled by $80\times 80$ cells. The ions are sampled by 125 macro-particles per cell. The electron fluid follows the adiabatic equation of state $T_e/n_e^{\gamma_e-1}=\rm const$, where $n_e$ is the electron density and $\gamma_e=5/3$.  
The time step is $\Delta t=0.01 \Omega_{ci}^{-1}$ where $\Omega_{ci}$ is the ion cyclotron frequency. 


The injection fields of a finite-scale, left-hand circularly polarized Alfv\'en wave are prepared as follows. 
First, in Simulation~\#1, a linearly polarized $B_x$ field is prescribed at $z=35 d_i$ with the distribution $B_x(x,y)=A\cos[\pi(x-x_0)/2r_s]\cos[\pi(y-y_0)/2r_s]\cos(\omega_0t)$ for $r=\sqrt{(x-x_0)^2+(y-y_0)^2}\leq r_s$ and $B_x(x,y)=0$ otherwise, where $A\ll 1$ (to avoid nonlinear interactions), $r_s=L_x/8$, ($x_0, y_0$) represents the center of the perpendicular plane, and $\tilde{\omega}_0\equiv\omega_0/\Omega_{ci}=0.31$; then the downstream magnetic fields (both $x,\ y$ components) at the perpendicular plane $z=40 d_i,\ t=100 \Omega_{ci}^{-1}$ are extracted as the first set of base fields $[B_{x1}(x,y),\ B_{y1}(x,y)]$. 
Note that while $\nabla\cdot B_x(x,y)\neq 0$ in this preparation run, $[B_{x1}(x,y),\ B_{y1}(x,y)]$ taken downstream after processed by the field solver are divergence free.  
We repeat this process in Simulation~\#2, where we prescribe a linearly polarized $B_y$ field
and obtain the second set of base fields $[B_{x2}(x,y),\ B_{y2}(x,y)]$. 
For actual physics runs, we inject the Alfv\'en wave by prescribing at $z=35 d_i$ the following combined base fields: 
\begin{subequations}
\begin{align}
    \frac{B_x(x,y)}{B_0}&=\frac{\delta B_x}{B_0}\bigg[\frac{B_{x1}(x,y)}{B_{1}^{\rm max}}\cos(\omega_0 t)- \frac{B_{x2}(x,y)}{B_{2}^{\rm max}}\sin(\omega_0 t)\bigg],\\
    \frac{B_y(x,y)}{B_0}&=\frac{\delta B_y}{B_0}\bigg[\frac{B_{y1}(x,y)}{B_{1}^{\rm max}}\cos(\omega_0 t)- \frac{B_{y2}(x,y)}{B_{2}^{\rm max}}\sin(\omega_0 t)\bigg],
\end{align}
\end{subequations}
where $\delta B_x/B_0=\delta B_y/B_0\equiv\delta B/B_0$ is the normalized wave amplitude and 
$B_{1}^{\rm max},\ B_{2}^{\rm max}$, found at $(x_0, y_0)$, are the maximum value of $\sqrt{B_{x1}^2+B_{y1}^2},\ \sqrt{B_{x2}^2+B_{y2}^2}$, respectively.
The injection contains a small ring-up time of 50$\Omega_{ci}^{-1}$ and lasts for $3000 \Omega_{ci}^{-1}$ 
The dispersion relation of resulting Alfv\'en wave in the downstream was checked and verified. 

The transverse field patterns both at the injection and downstream
are displayed in Figs.~\ref{fig1}(d1-d4), where this simulation case uses parameters $\tilde{\omega}_0=0.63$, $T_e/T_i=9$, and total beta $\beta=\beta_e+\beta_i=1\times 10^{-3}$. As the finite-scale wave propagates essentially in an Alfv\'en wave cone~\cite{morales1997structure}, the wave pattern rotates and spans multiple cycles in the perpendicular plane, giving a dominant $k_\perp d_i\simeq 2.2$ (following a Bessel function fit~\cite{morales1997structure}). Figure~\ref{fig1}(a) shows the contour surface of the perpendicular wave field $B_\perp=\sqrt{B_x^2+B_y^2}$ (for an initial injection of amplitude $\delta B/B_0=0.01$) at $t=200\Omega_{ci}^{-1}$, well before the onset of PDI. This finite-scale wave contains two rotating parallel electric currents [Fig.~\ref{fig1}(b)].
The parallel currents exert an influence on both the wave fields and ion dynamics through a constant resistivity $\eta=4\pi \nu_{ei}/\omega_{pe}^2$ used in the hybrid code to mimic electron-ion collisions, where $\nu_{ei}$ is the collisional rate and $\omega_{pe}$ the electron plasma frequency; see the Ohm's law and ion motion equation of the hybrid system~\cite{winske1991hybrid}:
\begin{equation}\label{ohmslaw}
    E+\frac{u_i\times B}{c} = \eta J + \frac{1}{q_i n_ic}J\times B - \frac{1}{q_i n_i}\nabla P_e,
\end{equation}
\begin{equation}\label{ion_motion}
    m_i \frac{d v_i}{dt} = q_i (E+v_i\times B/c)-e\eta J,
\end{equation}
where $E$ is the electric field, $P_e$ is the electron pressure tensor, $J$ is the total plasma current, $e$ is the elementary charge, $c$ is the light speed in a vacuum, and ($u_i, v_i, q_i, m_i, n_i$) are the ensemble ion speed, individual ion speed, ion charge, ion mass, and ion density, respectively. In the present case, the normalized resistivity used in the code is $\tilde{\eta}=\eta\omega_{pi}/4\pi=1\times 10^{-5}$ ($\omega_{pi}$ is the ion plasma frequency). 

\begin{figure*}[htp]
	\centering
	\includegraphics[width=0.9\textwidth]{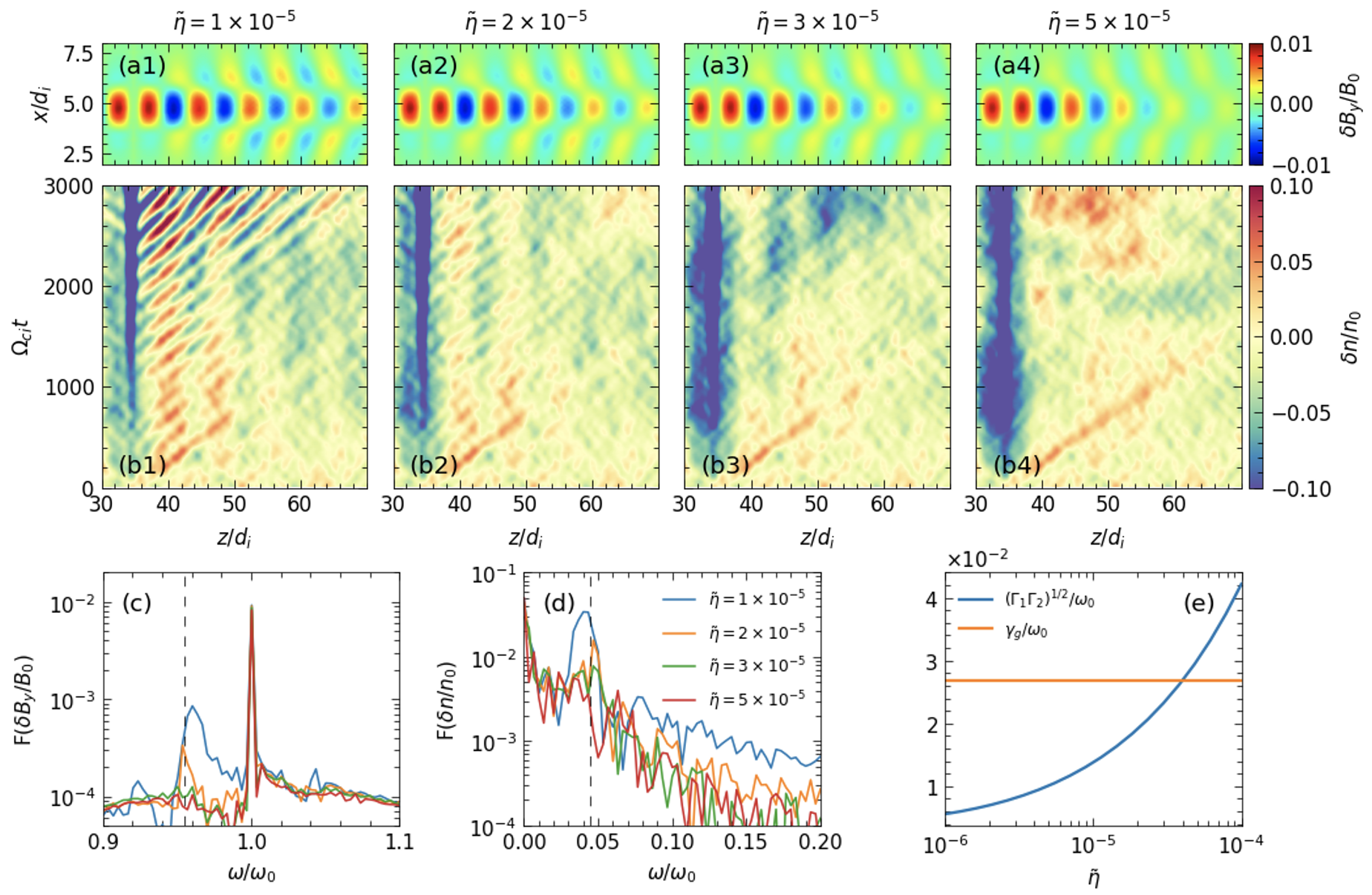}
	\caption{Effects of wave damping on PDI, while keeping the perpendicular wave scale $k_\perp d_i$ fixed to 2.2. The simulations correspond to the same beta $\beta=5\times 10^{-4}$, $T_e/T_i=4$, and $\omega_0/\Omega_{ci}=0.63$, with the resistivity shown in the titles. Panel (a) shows the central $xz$-cuts of the wave field component $\delta B_y/B_0$. Panel (b) shows the space-time evolution of the density fluctuation on the central axis $(x,y)=(5,5)d_i$. (c,d) The Fourier spectra of $\delta B_y/B_0$  and density fluctuations $\delta n/n_0$ probed at $(x,y,z)=(5,5,41) d_i$ for all four cases. (e) Theoretical calculation of the damping rate geometrical mean $(\Gamma_1\Gamma_2)^{1/2}/\omega_0$ versus varying $\tilde{\eta}$ and its comparison with the growth rate $\gamma_g/\omega_0$. }
	\label{fig2}
\end{figure*}

The evidence of PDI is partly illustrated by the density fluctuations shown at $t=2000\Omega_{ci}^{-1}$  [Fig.~\ref{fig1}(c)], after the instability has sufficiently developed. The fluctuations are associated with the child acoustic wave, which co-propagates with the pump Alfv\'en wave. The acoustic wave develops bowed isosurfaces as a result of nonlinear frequency shifts, i.e. central axis corresponds to larger wave amplitudes and more frequency shifts toward the smaller end. To further confirm the PDI signatures, we probe the temporal evolution of the field envelope $|\delta B_y/B_0|$ and ion charge-density fluctuations $\delta n/n_0$ at a fixed location and display the result in Fig.~\ref{fig1}(e). These fluctuations start to emerge after a few hundreds of $\Omega_{ci}^{-1}$ and continue to increase throughout the simulation. The $B$ field envelope oscillates due to the pump ($\omega_0$) beating with the child Alfv\'en wave ($\omega_1$) with a frequency difference $\Delta\omega =\omega_0-\omega_1$, and the density fluctuation oscillates at the eigen acoustic wave frequency $\omega$. Their growth matches each other in time, and their similar oscillation frequency $\Delta\omega\simeq \omega$ verifies the frequency matching condition $\omega_0=\omega_1+\omega$ as required for PDI coupling~\cite{sagdeev1969nonlinear,derby1978modulational,goldstein1978instability}. The frequency matching is also revealed in the spectra, Figs.~\ref{fig1}(f, g), where the predicted frequencies of the child waves are indicated by the vertical dashed lines. The excitation is so strong in this case that PDI coupling with harmonics of the acoustic mode $N\omega$ ($N$ is an integer) is also visible.

\section{Effects of 3D wave damping} \label{wave_damping}

The more realistic 3D finite-scale injection involves two new features (compared to a plane wave injection~\cite{li2022hybrid}): i) The Alfv\'en wave itself suffers from damping, 
including resistive damping [see Eq.~(\ref{ohmslaw})] and geometrical attenuation induced equivalent damping~\cite{morales1997structure}. 
ii) The finite $k_\perp$ associated with the finite perpendicular scale. 
While the wave damping is also dependent on $k_\perp$ (as we will see more clearly later), we will explore the effects of these two new features separately. 
Isolating the effects of $k_\perp$ is of interest because $k_\perp$ is potentially an important parameter determining PDI growth~\cite{matteini2010parametric,shi2019three}.    

We first examine the effects of wave damping on PDI excitation using a set of simulations with the outcome summarized in Fig.~\ref{fig2}. The magnitude of wave damping in these runs is controlled by the constant resistivity $\tilde{\eta}$, while the wave $k_\perp$ is fixed as the cases have the same perpendicular dimensions and same source size as used for Fig.~\ref{fig1} ($L_x,L_y=10d_i,\ r_s=L_x/8=1.25d_i$). The total beta is $\beta=5\times 10^{-4}$ with $T_e/T_i=4$. For each run, we display a snapshot of central $xz$-cut of the wave field component $\delta B_y/B_0$ at $t=200\Omega_{ci}^{-1}$ (before PDI develops) in Fig.~\ref{fig2}(a) and the space-time evolution of on-axis density fluctuations till the end of the simulation $t_{\rm max}=3000 \Omega_{ci}^{-1}$ in Fig.~\ref{fig2}(b). It is seen that by increasing $\tilde{\eta}$ from $1\times 10^{-5}$ to $5\times 10^{-5}$, the damping of the Alfv\'en wave is indeed much enhanced. While the density fluctuation (evidence of PDI) is strong for $\tilde{\eta}=1\times 10^{-5}$, it becomes much weaker for $\tilde{\eta}=2\times 10^{-5}$ and nearly disappeared when $\tilde{\eta}\geq 3\times 10^{-5}$. Figures 2(c,d) show more quantitative evidence of PDI by looking at the probe data (as done for Fig.~\ref{fig1}); similarly, prominent spectral peaks for the child Alfv\'en wave [Fig.~\ref{fig2}(c)] and child acoustic mode [Fig.~\ref{fig2}(c)] are found only for $\tilde{\eta}\leq 2\times 10^{-5}$. 

To physically and more quantitatively understand the wave damping effects, we estimate the damping rate for each damping mechanism. The resistive Alfv\'en wave damping essentially comes from the damping of the current channels through electron-ion collisions (represented by $\eta$). The total magnetic field evolves according to $\frac{\partial B}{\partial t} = \nabla\times(u_i\times B) - \frac{c^2}{4\pi}\eta \nabla\times(\nabla\times B) +\frac{c}{4\pi q_i}\nabla\times [\frac{1}{n_i}B\times (\nabla\times B)]$, obtained by taking the curl of  Eq.~(\ref{ohmslaw}) and making use of Faraday's law and Ampere's law. The first term on the right-hand side denotes motion of field lines frozen-in to the plasma, the second term denotes the resistivity-induced diffusion with the diffusion rate $D_r=\frac{c^2}{4\pi}\eta k^2$, the third term is the Hall term, and the electron pressure term is dropped as we expect an isotropic pressure due to collisions. A two-fluid analysis~\cite{mallet2023nonlinear} finds that the resistive damping rate is related to the diffusion rate as $\Gamma_r=\frac{1}{2(1+k_z^2d_i^2+k^2d_e^2)}D_r=\frac{1}{2(1+k_z^2d_i^2)}D_r$, where $kd_e\to 0$ for our massless electron fluid.   
By normalizing to the pump wave frequency, the resistive damping rate can be cast as
\begin{equation} \label{alfvendampingrate}
    \frac{\Gamma_r}{\omega_0}=\frac{1}{2}\frac{1+k_\perp^2/k_z^2}{1+k_\perp^2\rho_s^2}\frac{\omega_{pi}}{\Omega_{ci}}\tilde{\omega}_0\tilde{\eta},
\end{equation}
where $\rho_s=c_s/\Omega_{ci}$ is the ion sound gyroradius, $c_s=\sqrt{\frac{T_e+T_i}{m_i}}$ is the sound speed, and we have used the dispersion relation for a finite-frequency, finite-scale kinetic Alfv\'en wave~\cite{stasiewicz2000small} (neglecting the finite Larmor radius effects)
\begin{equation} \label{kaw-dispersino}
    \omega_0=k_z v_A\sqrt{1-\tilde{\omega}_0^2+k_\perp^2\rho_s^2}\equiv k_z v_A\sqrt{\Xi}.
\end{equation}

The equivalent Alfv\'en wave damping associated with geometrical attenuation may be estimated as follows. Let us define the effective wave source radius as $R=\sqrt{\frac{\iint (\delta B_\perp)^2 r^2dxdy}{\iint (\delta B_\perp)^2 dxdy}}$. The effective radius increases by $dR$ in a propagation distance of $dz$, according to the Alfv\'en wave cone angle 
$\tan\theta=\frac{dR}{dz}=\frac{v_{g,\perp}}{v_{g,z}}=\frac{\partial k_z}{\partial k_\perp}=\sqrt{\frac{\beta}{2}}\frac{k_\perp\rho_s}{\Xi^{3/2}}\tilde{\omega}_0$ \cite{morales1997structure}
where $\beta=2(c_s/v_A)^2$ and $v_{g,\perp}=\partial \omega_0/\partial k_\perp$, $v_{g,z}=\partial\omega_0/\partial k_z$ are the perpendicular and parallel group velocity, respectively. Without wave dissipation, the wave energy conservation at the two locations requires $(\delta B_1)^2 R^2 = (\delta B_2)^2(R+dR)^2$, which gives 
$\frac{\delta B_2}{\delta B_1}\simeq 1-dR/R=1-\frac{\tan\theta}{R}dz$. 
Comparing $\frac{\delta B_2}{\delta B_1}$ with an exponential on-axis wave damping/reduction due to geometrical spreading $\delta B=(\delta B)_0\exp(-S_g dz)\simeq (\delta B)_0 (1-S_g dz)$, one has $S_g = \frac{\tan\theta}{R}\simeq 0.25 k_\perp\tan\theta$; here we have used $k_\perp\simeq 4/R$ which is verified against simulations for different wave source sizes. Converting $S_g$ to the temporal damping rate gives
\begin{equation}\label{geometrical_damping}
    \frac{\Gamma_g}{\omega_0} = \frac{S_g v_{g,z}}{\omega_0}=0.18\frac{k_\perp}{k_z}\frac{k_\perp\rho_s}{\Xi^{3/2}+\tilde{\omega}_0^2\Xi^{1/2}}\tilde{\omega}_0\sqrt{\beta},
\end{equation} 
where $v_{g,z}=\frac{\omega_0/k_z}{1+\tilde{\omega}_0^2/\Xi}$.

Finally, the ion acoustic wave (or sound wave) Landau damping rate may be estimated as~\cite{li2022hybrid}
\begin{equation}\label{soundwave_damping}
    \frac{\Gamma_s}{\omega_0}\simeq 2\sqrt{\beta}\sqrt{\frac{T_i}{T_e}}.
\end{equation}
The resistivity also incurs a friction force $-e\eta J$ on the ion motion, as seen in Eq.~(\ref{ion_motion}). However, the friction force works through the current channels, which reside off the central axis. Therefore, the potential impact of the friction force on the damping of the sound wave peaked on the central axis [see Fig.~\ref{fig1}(c)] can be neglected. 

In a system with significant wave damping present, the PDI may be excited only if the following condition is satisfied~\cite{pesme1973parametric,montgomery2016two}:
\begin{equation}\label{PDI_threshod}
    \gamma_g/\omega_0>\sqrt{\Gamma_1 \Gamma_2}/\omega_0\equiv \Gamma_{gm}/\omega_0,    
\end{equation}
where $\gamma_g\simeq \frac{1}{2}(\delta B/B_0)/\beta^{1/4}$ is the PDI growth rate obtained for $k_\perp=0$ and $\Gamma_{gm}$ is the damping rate geometrical mean constructed from the damping rates of the two child modes: $\Gamma_1=\Gamma_r+\Gamma_g,\ \Gamma_2=\Gamma_s$. The use of zero-$k_\perp$ growth rate will be justified later where PDI excitation shows no discernible dependence on $k_\perp$ alone. Additionally, we essentially use the pump wave damping to approximate child Alfv\'en wave damping, because the two waves have the same nature except for a minor frequency difference $\sim 2\sqrt{\beta}\omega_0$ at low beta. To test Eq.~(\ref{PDI_threshod}), we substitute the common simulation parameters $\omega_0/\Omega_{ci}=0.63$, $\omega_{pi}/\Omega_{ci}=300$, $\beta=5\times10^{-4}$, $T_e/T_i=4$ and $k_\perp d_i=2.2$ into $(\Gamma_r, \Gamma_g, \Gamma_s, \gamma_g)$. We take the wave amplitude $\delta B/B_0=0.8\times 10^{-2}$ at $z=41 d_i$ (close to the injection, where PDI is probed) for the growth rate calculation; if PDI cannot be excited close to the injection, it cannot be excited in the rest of the domain where the wave amplitudes are smaller. The result shows that $\Gamma_g/\omega_0\simeq 1.8\times 10^{-4}$, $\Gamma_s/\omega_0\simeq 2.2\times 10^{-2}$, and $\gamma_g/\omega_0\simeq 2.7\times 10^{-2}$, i.e. $\gamma_g>\Gamma_s\gg \Gamma_g$. Therefore, whether PDI can be excited is strongly dependent on the resistive damping $\Gamma_r/\omega_0$. By varying $\tilde{\eta}$ (hence $\Gamma_r/\omega_0$), the two sides of Eq.~(\ref{PDI_threshod}) plotted in Fig.~\ref{fig2}(e) shows that Eq.~(\ref{PDI_threshod}) is satisfied only for $\tilde{\eta}<4\times 10^{-5}$. The good agreement with the 3D simulations confirms our physical understanding and underscores the importance of wave damping for PDI excitation by a 3D finite-scale kinetic Alfv\'en wave. Notice that Eq.~(\ref{PDI_threshod}) necessarily modifies the threshold Alfv\'en wave amplitude obtained in the plane-wave study~\cite{li2022hybrid}, due to the new 3D wave damping presented here.


\section{Effects of finite perpendicular scale} \label{wave_kperp}

\begin{figure*}[htp]
	\centering
	\includegraphics[width=0.9\textwidth]{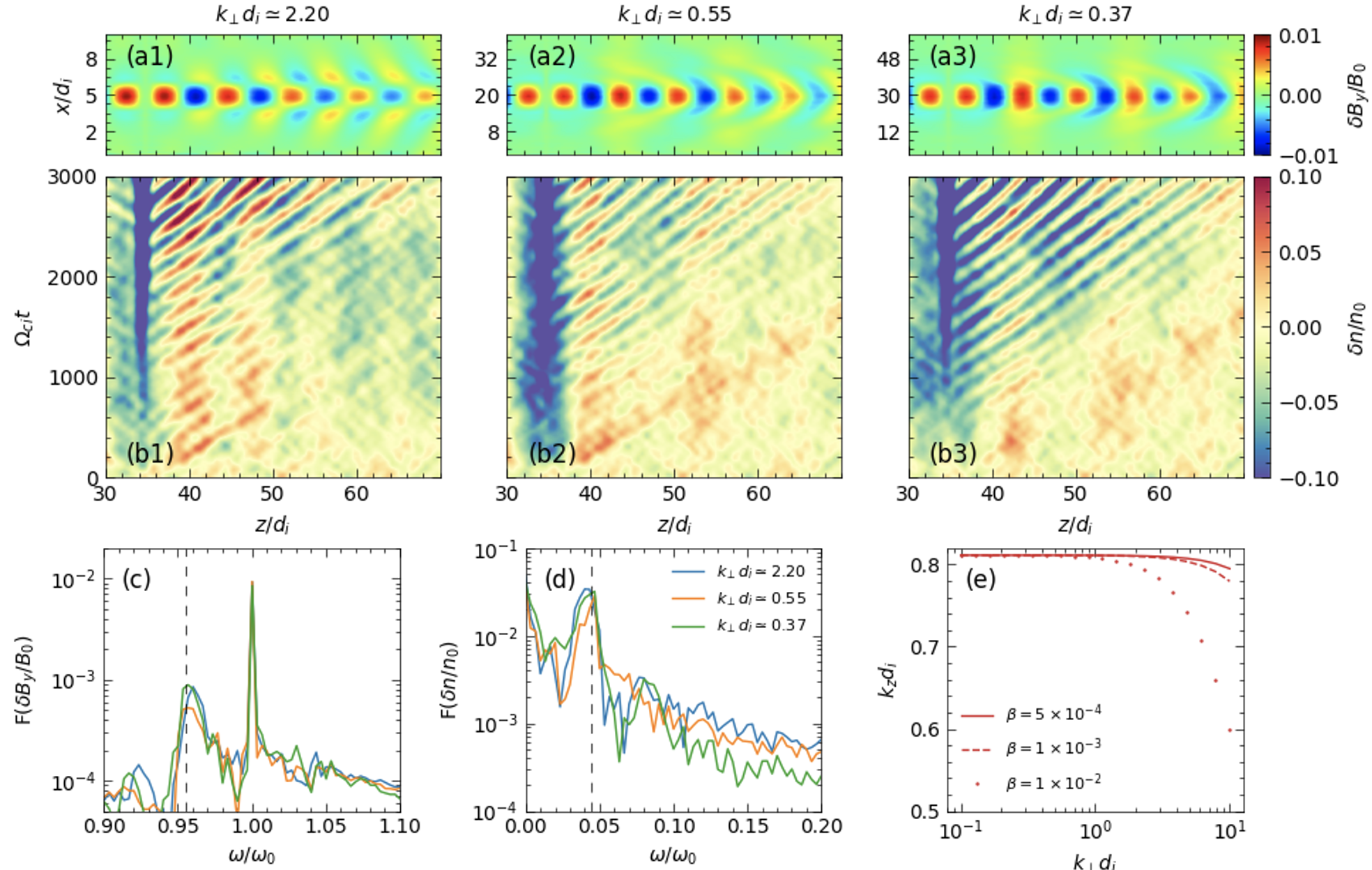}
	\caption{Effects of perpendicular wave scale on PDI excitation under constant wave damping. The simulations correspond to the same beta $\beta=5\times 10^{-4}$ and $T_e/T_i=4$ but varying perpendicular wave scales (see the titles). To maintain constant damping, the normalized resistivity for the cases shown from left to right is adjusted to be $\tilde{\eta}=1, 5, 6\times 10^{-5}$, respectively. Panels (a-d) have the same format with Fig.~\ref{fig2}. 
    (e) Dependence of $k_zd_i$ on the perpendicular wave scale for a kinetic Alfv\'en wave, calculated for different $\beta$ and same $T_e/T_i=4,\ \omega_0/\Omega_{ci}=0.63$. 
    }
	\label{fig3}
\end{figure*}

We next vary $k_\perp$ to explore the effects of finite perpendicular wave scale on PDI excitation under constant wave damping. Since the resistive damping [Eq.~(\ref{alfvendampingrate})] and geometrical attenuation [Eq.~(\ref{geometrical_damping})] also change with $k_\perp$, we simultaneously adjust $\tilde{\eta}$ in the simulations to keep $\Gamma_{gm}$ close to a constant. Figure 3 displays the outcome of three cases satisfying the above requirements, where the normalized resistivity for cases of different wave source sizes ($r_s = 1.25 d_i,\ 5d_i,\ 7.5 d_i$) is adjusted to be $\tilde{\eta}=1, 5, 6\times 10^{-5}$, respectively. With these parameters and their common setup $\delta B/B_0\simeq 0.8\times 10^{-2}$ (at the probe location), $\omega_0/\Omega_{ci}=0.63$, and $\omega_{pi}/\Omega_{ci}=c/v_A=300$, the Alfv\'en wave damping for the three cases, the Alfv\'en wave damping $(\Gamma_r+\Gamma_g)/\omega_0$ is kept at about 0.007. As shown in Figs.~\ref{fig3}(a1-a3), the wave magnetic fields $\delta B_x/B_0$ of these cases indeed have a similar spatial wave damping profile. 
The sound wave damping, $\Gamma_s/\omega_0\simeq 0.022$, has no dependence on $k_\perp$. Therefore, the wave damping geometrical mean for all three cases is $\Gamma_{gm}/\omega_0\simeq 0.012$. The simulation results shown in Fig.~\ref{fig3} reveal little difference in PDI excitation, both in terms of the space-time evolution of on-axis density fluctuations and probed spectra, despite the factor of six difference in $k_\perp d_i$ among the three cases. This result strongly suggests that the PDI excitation has no discernible dependence on the perpendicular wave scale alone. 

The result may seem to contradict some 1D/2D periodic-boundary simulations of a finite-$k_\perp$ plane Alfv\'en wave reported previously~\cite{matteini2010parametric}, where a $\cos\theta_{kB}$ dependence of PDI growth rate was extracted ($\theta_{kB}=\arctan(k_\perp/k_z)$ is the normal angle of the oblique Alfv\'en wave). The three cases shown in Fig.~\ref{fig3} have $k_\perp/k_z \simeq 2.71, 0.68, 0.45$, corresponding to a wave normal angle of 70, 34, 24 degrees and $\cos\theta_{kB}$ of 0.35, 0.83, 0.91, respectively, which span a variation large enough to discern the potential consequences of the $\cos\theta_{kB}$ dependence. The apparent contradiction may be understood as follows: The PDI growth rate crucially depends on the parallel ponderomotive force $F_{p,\parallel}$ that drives the acoustic mode.  In the plane-wave scenario of~\cite{matteini2010parametric}, the Alfv\'en wave was loaded with wavelength $\lambda_0$ along $z$, and $\theta_{kB}$ was introduced by tuning the $B_0$ direction away from $z$; as a result, the parallel wavelength along the background field becomes $\lambda_\parallel=\lambda_0/\cos\theta_{kB}$, the perpendicular wave electric field is $\delta E_\perp=\delta E\cos\theta_{kB}$, and the wave frequency $\omega=k_\parallel v_A$ is $\cos\theta_{kB}$ times smaller. This results in a parallel ponderomotive force $F_{p,\parallel}\propto \omega_0^{-2}\nabla_\parallel (\delta E_\perp)^2$ which is $\cos\theta_{kB}$ times smaller. By contrast, in the finite-scale scenario considered in the present work, the parallel wavelength or $k_z$ changes with the source scale according to the dispersion relation Eq.~(\ref{kaw-dispersino}), which can be recast as
\begin{equation}
k_z d_i=\tilde{\omega}_0/\sqrt{1-\tilde{\omega}_0^2+\frac{1}{2}k_\perp^2d_i^2\beta},    
\end{equation}
where we have used $k_\perp^2\rho_s^2=\frac{1}{2}k_\perp^2 d_i^2\beta$. A calculation of $k_zd_i$ versus a broad range of perpendicular wave scales under different $\beta$ is shown in Fig.~\ref{fig3}(e). It is seen that, despite the large $k_\perp$ or $\theta_{kB}$, the parallel wavelength depends only weakly on the source size, especially for the low-beta regime with $\beta<10^{-2}$. Meanwhile, the driving frequency is fixed and the perpendicular wave field $\delta B_\perp$ remains the same as the cases in Fig.~\ref{fig3} are tuned to have similar wave damping. Therefore, the parallel ponderomotive force in our scenario is similar, leading to similar PDI growth when varying the perpendicular wave scale alone. 

\section{Discussion} \label{discussion}

These 3D simulations and associated analyses suggest that the wave damping is the major limiting factor for driving finite-scale Alfv\'en wave PDI in a uniform background, while $k_\perp$ alone shows no discernible influence. As illustrated by Eq.~(\ref{alfvendampingrate}), Eq.~(\ref{geometrical_damping}) and Eq.~(\ref{soundwave_damping}, the wave damping depends on multiple parameters, e.g. the electron-ion collisional rate $\nu_{ei}$, wave normal angle $\propto k_\perp/k_z$, driving frequency $\tilde{\omega}_0$, total plasma beta $\beta$, temperature ratio $T_e/T_i$, and the absolute Alfv\'en speed $v_A/c$. The explicit scaling versus these parameters will help extrapolate to a broad range of parameters with relevance to both the laboratory and space plasmas. 

The low-beta laboratory plasma usually involves highly oblique waves $k_\perp d_i\sim\mathcal{O}(10)$ and $T_e/T_i\gg1$, $\tilde{\omega}_0\lesssim 1$. 
With a typical set of LAPD parameters, $\delta B/B_0=0.7\times 10^{-2}$, $\tilde{\omega}_0=0.6$, $c/v_A=660$, $n_e=3.5\times 10^{12} \rm cm^{-3}$, $\beta=1.54\times 10^{-3}$, $T_e/T_i=5.5$, $k_\perp d_i=15.28$, $\tilde{\eta}=1\times 10^{-6}$ ($\nu_{ei}=8.6$ MHz, electron skin depth $d_e \simeq 3$ mm), and the ion species $\rm He^+$, the calculated resistive damping is $\Gamma_r/\omega_0\simeq 0.09$, geometrical attenuation rate $\Gamma_g/\omega_0\simeq 0.038$, sound wave damping $\Gamma_s/\omega_0\simeq 0.033$ and the damping rate geometrical mean $\Gamma_{gm}/\omega_0\simeq 0.065$, while the growth rate under this set of parameters is only $\gamma_g/\omega_0\simeq 0.018$. The damping rate is nearly four times larger, which may explain why PDI has been difficult to excite on LAPD using a single wave driver. Actual experiments also involve electron Landau damping of the Alfv\'en wave and ion-neutral collisional damping of the sound wave, which  
will further raise the damping rates and thereby constrain PDI excitation. 

To excite PDI essentially requires the PDI growth to overcome the wave damping.
While a smaller $k_\perp$ helps reduce Alfv\'en wave damping $\Gamma_r$ and $\Gamma_g$ significantly, a new, next-generation laboratory facility would be required to launch an Alfv{\'e}n wave with a much larger perpendicular size ($k_\perp d_i < 1)$ \cite{dorfman2023nextgen}. 
Yet our study shows that the PDI growth is not compromised by the large $k_\perp$.  
Therefore, future optimization strategies 
should focus on reducing wave damping for currently achievable values of $k_\perp$ by varying other parameters.  
One option is to achieve higher electron temperatures, while maintaining a cold ion population. Hotter electrons will reduce electron Landau damping and electron-ion collisional damping (e.g. $\eta$ in our simulations) of the Alfv\'en wave; a larger temperature ratio $T_e/T_i$ will help suppress the sound wave damping. Higher electron temperatures may also improve antenna-plasma coupling~\cite{gigliotti2009generation}, leading to larger driving wave amplitudes $\delta B/B_0$. 
To achieve PDI excitation in a bounded laboratory plasma, care must also be taken to operate at sufficiently high driving wave frequency~\cite{li2022hybrid}. Based on the 3D damping effects elucidated in this paper, 
it may be beneficial to operate at as low of a pump frequency as possible
to both minimize Alfv\'en wave damping 
and increase the parallel (along $z$) ponderomotive force, i.e. $F_{p,z}\propto \omega_0^{-2}\nabla_\parallel (\delta E_\perp)^2\propto \tilde{\omega}_0^{-2} k_z\propto \frac{1}{\tilde{\omega}_0\sqrt{1-\tilde{\omega}_0^2}}$ becomes larger at smaller $\tilde{\omega}_0$.

The low-beta space plasmas, on the other hand, have distinct wave and plasma properties from the laboratory (albeit similar dimensionless/scaled parameters, for example, in the solar coronal region~\cite{bose2019measured}), such as a very low frequency $\tilde{\omega}_0\ll 1$, comparable electron/ion temperatures $T_e/T_i\sim 1$, and a large wave amplitude $\delta B/B_0$~\cite{reville2018parametric,matteini2024alfvenic}.
Taking $\delta B/B_0=0.2$, $\tilde{\omega}_0=0.01$, $c/v_A=600$, $\beta=2\times 10^{-2}$, $T_e/T_i=1$, $k_\perp d_i=0.55$ and $\tilde{\eta}=1\times 10^{-8}$ for example, a calculation based on the present framework shows that the growth rate $\gamma_g/\omega_0\simeq 0.27$ is much larger than the damping rate geometrical mean $\Gamma_{gm}/\omega_0\simeq 0.016$, where $\Gamma_r/\omega_0\simeq 9\times 10^{-5}$, $\Gamma_g/\omega_0\simeq 8\times 10^{-4}$, and $\Gamma_s/\omega_0\simeq 0.28$. 
Therefore, it is potentially much easier to excite PDI in space plasma than in the laboratory. Interestingly, while $\Gamma_{gm}$ is small, a significant sound wave damping $\Gamma_s\sim\gamma_g\gg \Gamma_{gm}$ is allowed owing to the way $\Gamma_{gm}$ is constructed. In other words, PDI in this space-relevant case is mainly facilitated by the negligible Alfv\'en wave damping, while the appreciable sound wave damping facilitates final PDI energy dissipation. 
Certainly, the above estimates based on a uniform background may be complicated by the strong inhomogeneity present in space plasmas. In particular, the Alfv\'en wave in the solar coronal region may develop large $k_\perp$~\cite{ofman1995alfven,hollweg1999kinetic,kiyani2015dissipation} among other complications. 
The independence of PDI versus $k_\perp$ alone, as discovered in this work, will be of key importance to establishing the relevance of PDI at small dissipation scales. 

\section{Summary}\label{summary}

In summary, we have presented the first 3D open-boundary hybrid simulations of PDI driven by a single Alfv\'en wave of finite perpendicular scale.  
It is found that the PDI excitation 
is strongly limited by 3D wave damping,
while the perpendicular wave scale ($k_\perp$) alone plays no discernible influence on PDI. 
These results are crucial to understanding the excitation criteria of Alfv\'en wave PDI in a practical 3D open system. In the laboratory with very small-scale waves, PDI is mainly hindered by the strong wave damping. 
Strategies to optimize experimental parameters to minimize the damping rates and enhance PDI growth rate are briefly discussed, which will be essential for demonstrating PDI in future laboratory experiments. 
In low-beta space plasmas, PDI excitation will benefit greatly from our finding that the PDI growth rate has no discernible dependence on $k_\perp$ alone, and the Alfv\'en wave properties in space are more likely to make PDI a relevant and important scheme leading to wave dissipation at both large and small scales. 
Future studies would involve elaborating on the dynamics/consequences of finite-scale Alfv\'en wave PDI with the 3D open system, as well as developing more comprehensive simulation models to include 
both electron Landau damping and ion-neutral collisional damping that are absent in the current hybrid code. It would also be important to investigate the effects of a nonuniform background (both in the perpendicular and parallel directions) which is commonly found in space plasmas.

\section{Acknowledgement}
This work was supported by the DOE grant DE-SC0021237 through the NSF/DOE Partnership in Basic Plasma Science and Engineering program, the DOE grant DE-SC0023893, and the NASA grant 80NSSC23K0695. 
We acknowledge the Texas Advanced Computing Center (TACC) at The University of Texas at Austin 
and the National Energy Research Scientific Computing Center (NERSC) 
for providing the computing and visualization resources. 

\section{Data availability statement}
The data that support the findings of this study are available from the corresponding author upon reasonable request.

\bibliography{ref_pdi}

\end{document}